\documentclass{article}
\usepackage[utf8]{inputenc}
\usepackage{graphicx}
\usepackage[a4paper, margin=1in]{geometry}
\usepackage[utf8]{inputenc}
\usepackage[T1]{fontenc}
\usepackage{microtype}
\usepackage[authoryear]{natbib}
\usepackage[hyphens]{url}
\usepackage{color}
\definecolor{darkblue}{rgb}{0.0, 0.0, 0.55}
\usepackage[colorlinks=true,linkcolor=darkblue,citecolor=darkblue,urlcolor=darkblue]{hyperref} 
\usepackage{comment}
\usepackage{tabularx}
\usepackage{booktabs}
\usepackage{longtable}
\usepackage{pdflscape}
\usepackage{geometry}
\usepackage{comment}
\usepackage{multirow}
\usepackage{array}
\usepackage[utf8]{inputenc}

\newcommand{\text}[1]{\mathrm{#1}}

\usepackage{caption}
\usepackage{libertine}
\usepackage{inconsolata}

\title{Beyond Static Responses: Multi-Agent LLM Systems as a New Paradigm for Social Science Research}

\author{
    Jennifer Haase\\
    Weizenbaum Institute and HU\\ Berlin, Germany\\
	\url{jennifer.haase@hu-berlin.de}
    \and
    Sebastian Pokutta\\
    TU Berlin and Zuse Institute Berlin \\ Berlin, Germany\\
    \url{pokutta@zib.de}
}

\date{June 2, 2025}

\begin{document}

\maketitle

\begin{abstract}
As large language models (LLMs) transition from static tools to fully agentic systems, their potential for transforming social science research has become increasingly evident. This paper introduces a structured framework for understanding the diverse applications of LLM-based agents, ranging from simple data processors to complex, multi-agent systems capable of simulating emergent social dynamics. By mapping this developmental continuum across six levels, the paper clarifies the technical and methodological boundaries between different agentic architectures, providing a comprehensive overview of current capabilities and future potential. It highlights how lower-tier systems streamline conventional tasks like text classification and data annotation, while higher-tier systems enable novel forms of inquiry, including the study of group dynamics, norm formation, and large-scale social processes. However, these advancements also introduce significant challenges, including issues of reproducibility, ethical oversight, and the risk of emergent biases. The paper critically examines these concerns, emphasizing the need for robust validation protocols, interdisciplinary collaboration, and standardized evaluation metrics. It argues that while LLM-based agents hold transformative potential for the social sciences, realizing this promise will require careful, context-sensitive deployment and ongoing methodological refinement. The paper concludes with a call for future research that balances technical innovation with ethical responsibility, encouraging the development of agentic systems that not only replicate but also extend the frontiers of social science, offering new insights into the complexities of human behavior.

\end{abstract}

\subsubsection*{A note to the reader} The field of AI---and especially agentic systems---is evolving at a rapid pace, with no expectation of reaching a final, stable state in the near future. Accordingly, this paper is intended as a ``living document'' that summarizes current trends and approaches. We plan to update it regularly as we receive feedback, as new developments emerge, as additional works are published, and as we continue to refine the framework.

If you need a stable identifier for the latest version of this paper please use the arxiv base url, e.g., \url{https://arxiv.org/abs/2505.xxxx}; for a specific version please use the versioned arxiv url, e.g., \url{https://arxiv.org/abs/2505.xxxxv1}. 

\section{Introduction}

Large language models (LLMs) have become foundational tools in contemporary social science research, widely applied to tasks such as survey response generation, qualitative analysis, coding, and summarization \citep{kantorBestPracticesImplementing2024, hardyLargeLanguageModels2023, demszkyUsingLargeLanguage2023}. These applications offer notable methodological benefits by replicating specific cognitive functions traditionally performed by human researchers or participants \citep{keExploringFrontiersLLMs2024, manningAutomatedSocialScience2024}. However, they remain largely confined to static and reactive uses, where models act as sophisticated text processors without memory, autonomy, or interactive agency---a limitation that echoes longstanding critiques of agentless computational modeling \citep{epsteinAgentbasedComputationalModels1999,epsteinGenerativeSocialScience2012}.

We argue that the next major step in computational social science lies in moving beyond these static applications toward a systematic understanding of LLM-based agentic systems \citep{fengWhenOneLLM2025,grossmannAITransformationSocial2023}. These systems differ fundamentally from conventional LLM use: they incorporate memory, goal-directed behavior, environmental interaction, and in some cases adaptive learning. Embedded within such architectures, LLMs do not merely mimic isolated cognitive acts---they function as interactive social agents capable of participating in simulations, decision-making, and complex group dynamics \citep{piaoAgentSocietyLargeScaleSimulation2025, wangSurveyLargeLanguage2024, luAIScientistFully2024}. Building on this foundation, we propose a six-tier framework that captures the increasing complexity and autonomy of LLM-based systems in social science research. This framework spans a conceptual continuum from static tools to fully agentic systems and is structured by functional thresholds---such as memory integration, autonomy, coordination, and learning---that define the degree of agentic behavior. At the foundational level, \textbf{LLM-as-Tool} systems (Level 0) operate as stateless text generators, producing contextually appropriate responses without memory, autonomy, or strategic reasoning. These systems rely solely on prompt inputs to generate outputs, lacking the capacity for long-term planning or adaptive behavior. Despite their limitations, they have proven effective for tasks like content generation, data classification, and text summarization, where immediate, context-free processing suffices \citep{thapaLargeLanguageModels2025, haaseApproachModelForgetting2020, karjusMachineassistedQuantitizingDesigns2025,ziemsCanLargeLanguage2024}. Moving to \textbf{LLM-as-Role} systems (Level 1), these architectures introduce basic state retention, enabling agents to simulate consistent personas or role-based behavior across multiple interactions. This level adds a layer of memory, allowing agents to maintain contextual awareness over short dialogue sequences, supporting applications like customer service, personalized tutoring, and basic psychological profiling \citep{wangEvaluatingAbilityLarge2025,sunRandomSiliconSampling2024}. However, these systems still lack the goal-directed autonomy required for independent decision-making. At \textbf{Agent-like LLM} (Level 2), systems gain more structured autonomy, integrating task-oriented reasoning and memory architectures that enable multi-step planning. These agents are capable of decomposing complex tasks, setting intermediate goals, and adjusting their behavior based on task outcomes \citep{hortonLargeLanguageModels2023,argyleOutOneMany2023}. They bridge the gap between static tools and fully autonomous agents, simulating more sophisticated human-like behavior, including context-driven decision-making and reflective judgment. The next tier, \textbf{LLM-based Agents} (Level 3), marks a significant step toward actual agency. These systems integrate comprehensive memory, environment interfaces, and strategic coordination mechanisms. They exhibit proactive behavior, leveraging long-term memory and environmental feedback to refine their actions over time. This level represents a crucial threshold in agentic complexity, where agents transition from passive responders to active decision-makers capable of planning, coordination, and strategic interaction \citep{manningAutomatedSocialScience2024,liuTwoHeadsAre2024}. \textbf{Multi-Agent Systems} (Level 4) extend this capability further, integrating multiple, interacting agents within a shared environment. These systems replicate complex social processes, such as negotiation, coalition-building, and organizational decision-making \citep{zhangExploringCollaborationMechanisms2024,fengWhenOneLLM2025}. They support distributed, collaborative problem-solving, with agents dynamically coordinating their actions based on shared goals and situational awareness. At this stage, agents exhibit emergent group behaviors, reflecting foundational principles from collective intelligence and network theory. Finally, at the highest level, \textbf{Complex Adaptive Systems} (Level 5) encompass large-scale, \textit{emergent} social dynamics. These architectures consist of numerous interacting agents, each equipped with memory, autonomy, and adaptive learning capabilities. Unlike lower tiers, these systems are characterized by self-organization, norm formation, and systemic adaptation, capturing the unpredictable, emergent properties of real-world social networks \citep{piaoAgentSocietyLargeScaleSimulation2025, wangInvestigatingExtendingHomans2025}. They provide powerful platforms for modeling phenomena like cultural evolution, institutional change, opinion-dynamics and large-scale social movements, pushing the boundaries of agentic LLM applications in computational social science.

These thresholds align with the OODA loop (Observe, Orient, Decide, Act; \citealt{boydDiscourseWinningLosing2018, osingaScienceStrategyWar2007}), providing a decision-theoretic lens to understand how LLM-based agents perceive, reason, and act in dynamic environments. This progression across the six tiers reflects the foundational structure of social science, moving from the study of individual cognition and behavior to the dynamics of small groups, and ultimately to the complex interactions of entire societies and populations. Grounding this structure in functional thresholds and aligning it with the OODA loop enables us to classify existing systems and clarify the developmental logic behind building LLM-based simulations. Through this lens, we present empirical examples and conceptual distinctions that show how such systems can explore social behavior, generate synthetic data, and simulate interactions at scale and ethical boundaries unattainable with human participants alone. By moving beyond narrow prompt-response paradigms toward dynamic, interactive, and adaptive architectures, this paper aims to lay a possible conceptual foundation for the next generation of computational social science.

\section{From Tools to Societies: A Framework of LLM-Based Agentic Systems}

The integration of LLMs into dynamic, agentic systems presents substantial potential for advancing computational social science. LLMs such as GPT-4 are already widely used for tasks like text generation, qualitative coding, and simulating survey responses \citep{gaoLargeLanguageModels2024, wangSurveyLargeLanguage2024}. However, these applications typically treat LLMs as static and reactive systems: they lack memory, do not interact with their environment, and pursue no autonomous goals. Their responses are shaped solely by the immediate prompt, without context retention beyond chat history, planning, or proactive behavior. While useful for emulating isolated cognitive functions (e.g., for personality modeling, \citealt{wangEvaluatingAbilityLarge2025, huangDesigningLLMAgentsPersonalities2024}; emotional expression, \citealt{mozikovGoodBadHulklike2024}; political personas, \citealt{vonderheydeAssessingBiasLLMGenerated2023}; or buyer preferences, \citealt{zhangGenerativeAgentsRecommendation2024}), these systems remain bounded by narrow task scopes and rigid interaction structures. In contrast, agentic systems represent a qualitatively different paradigm (cf. Table~\ref{tab:llm_agent_architecture}). They exhibit autonomous reasoning, context-aware memory, goal-directed behavior, and adaptive interaction with complex environments \citep{epsteinAgentbasedComputationalModels1999, epsteinGenerativeSocialScience2012,duenez-guzmanSocialPathHumanlike2023}. These systems maintain persistent internal states and memory, enabling them to draw on prior experiences for informed action \citep{huangDesigningLLMAgentsPersonalities2024}. Their autonomy is often grounded in predefined or evolving objectives, giving rise to dynamic, self-regulated responses across unfolding conditions \citep{fengWhenOneLLM2025, parkGenerativeAgentsInteractive2023}.

\begin{table}[t]
\centering
\renewcommand{\arraystretch}{1.15}
\small
\begin{tabularx}{\textwidth}{
  p{0.3cm} 
  >{\raggedright\arraybackslash}p{1.8cm} 
  >{\raggedright\arraybackslash}p{2cm} 
  >{\raggedright\arraybackslash}p{2.2cm} 
  >{\raggedright\arraybackslash}p{2.5cm} 
  >{\raggedright\arraybackslash}X}
\toprule
\textbf{Lvl} & \textbf{System Type} & \textbf{Threshold Criterion} & \textbf{OODA Phase} & \textbf{Key Capability} & \textbf{Required Architecture} \\
\midrule
0 & LLM-as-Tool & (baseline) & Act & Stateless response generation & LLM, prompt engineering \\
\addlinespace
1 & LLM-as-Role & Memory Integration & Observe → Act & Consistent persona or stateful behavior & Session memory, prompt engineering \\
\addlinespace
2 & Agent-like LLM & Goal-Driven Task Autonomy & Observe → Orient → Act & Purpose-driven action selection & Task objectives, control logic, multi-step prompts \\
\addlinespace
3 & Fully Agentic LLM & Environment Interface, Planning and Coordination & Observe → Orient → Decide → Act (OODA) & Interactive planning and strategic reasoning & Memory store, tool use, API access, coordination logic \\
\addlinespace
4 & Multi-Agent System & Multi-agent Coordination & OODA + Learning & Task division, negotiation, shared goals & Orchestration layer, agent communication protocols, shared memory or blackboard systems, role differentiation \\
\addlinespace
5 & Complex Adaptive System & Emergence + Adaptation & Dynamic OODA + Learning + Emergence & Population-level dynamics, norm formation, diffusion & Decentralized architecture, large-scale agent population, dynamic environment interfaces, feedback loops, adaptive learning and evolution mechanisms \\
\bottomrule
\end{tabularx}
\caption{LLM-Agentic System Continuum: Thresholds, OODA Mapping, and Architectural Requirements}
\label{tab:llm_agent_architecture}
\end{table}

Between these two extremes lies a conceptual middle ground: LLM-based, agent-like systems. These architectures exceed simple prompt-response configurations by integrating limited capabilities such as session memory, elementary tool use, or task-specific control logic. However, they fall short of full agentic autonomy and adaptability, lacking both sophisticated environmental interaction and long-term planning. The importance of this intermediate stage lies in its role as a stepping stone toward agentic complexity. It reflects a developmental transition in system design that is increasingly visible in practical applications. Recent work also underscores the potential of LLMs to serve as cognitive models, capturing aspects of human reasoning and aligning with foundational theories in cognitive science \citep{niuLargeLanguageModels2024}. This positions LLMs not only as functional tools but as plausible proxies for human-like agency, reinforcing their value in simulating social phenomena.

\begin{figure}[h]
    \centering
    \includegraphics[width=0.99\linewidth]{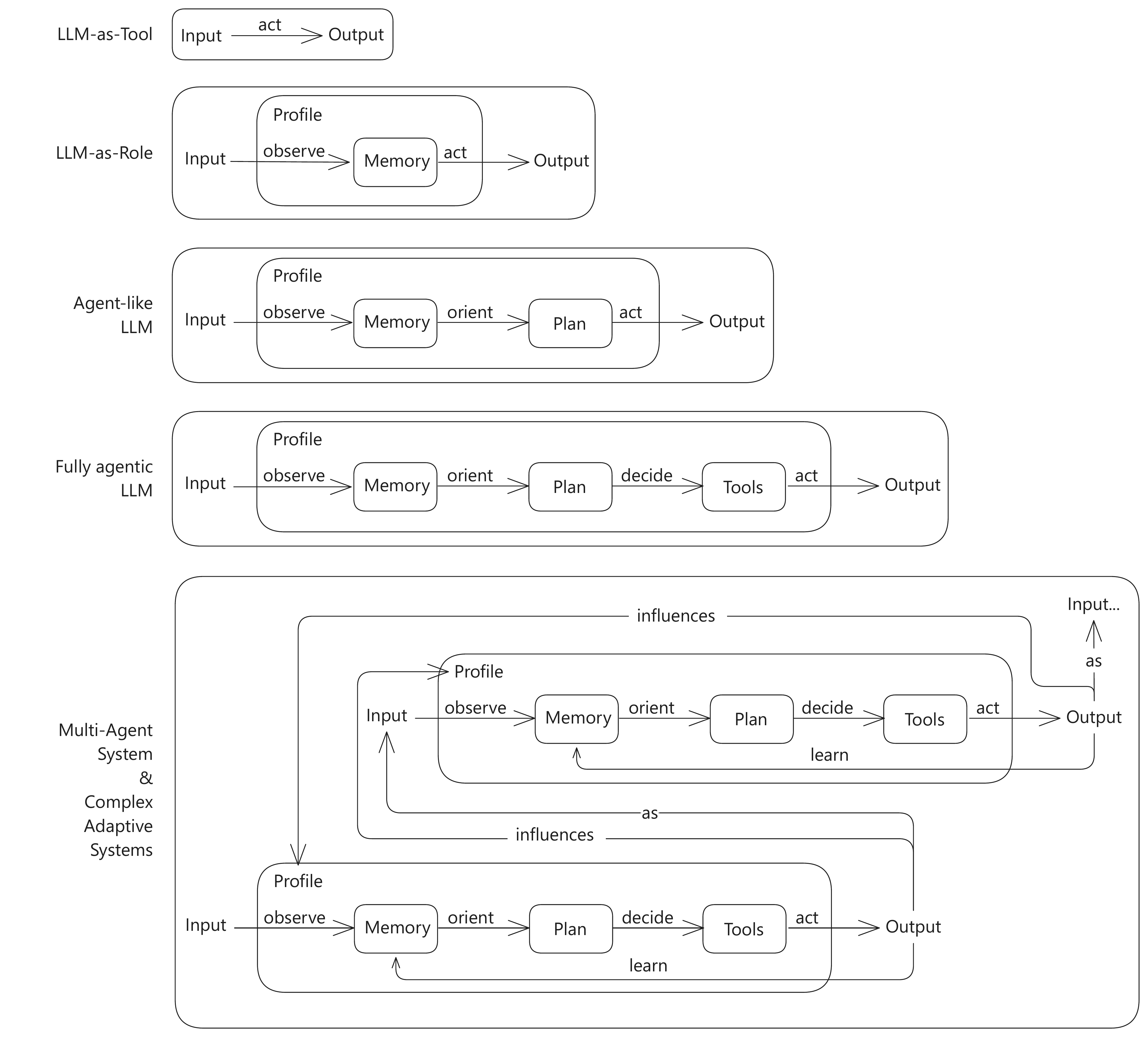}
    \caption{Design Architecture of the Levels of Agentic LLM Systems}
    \label{fig:level_agent}
\end{figure}

\subsection{Functional Thresholds for LLM-based Agents}
To systematically distinguish between these levels of system autonomy and interaction, we propose a continuum structured by a series of functional thresholds. These thresholds---memory integration, autonomy, planning and coordination, and adaptive learning---serve as markers of increasing agentic potential. Importantly, each threshold can be mapped to a deeper level of participation in the OODA loop (Observe, Orient, Decide, Act), a well-established model for adaptive decision-making in dynamic environments \citep{boydDiscourseWinningLosing2018,osingaScienceStrategyWar2007}. Initially developed by military strategist John Boyd, the OODA loop describes how intelligent agents continuously engage with their environment through a cyclical process: they observe their surroundings, orient themselves by interpreting new information in context, decide on a course of action, and then act—feeding the results back into the next observational phase. This loop is not a linear sequence but a recursive, feedback-driven mechanism central to real-time sense-making, learning, and adaptation. In the context of LLM-based systems, the OODA loop provides a functional scaffold for understanding different levels of agentic behavior (cf. Figure~\ref{fig:level_agent}). Static LLMs may only participate in the final ``act'' phase, producing text in response to prompts based on the LLM's innate capabilities. However, as these systems acquire new capabilities, such as memory, goal orientation, planning, and adaptive learning, they progressively engage in more complex phases of the loop---each level adds an OODA element. They begin to ``observe'' the inputs from the perspective of the given profile, which they can retain thanks to the memory (cf. Figure~\ref{fig:level_agent}, \textit{LLM-as-Role}). Thus, they can maintain a coherent persona, including personality or relatively stable opinions and values \citep{wangEvaluatingAbilityLarge2025,huangDesigningLLMAgentsPersonalities2024}. 

When adding a planning tool, LLMs can start to orient the output and ``behavior'' by updating internal states, deciding among competing options, and acting based on multi-step strategies---such tools are \textit{agent-like LLM}. To become fully agentic, they progressively gain the capacity to use tools for interactive, feedback-driven behaviors (cf. Figure~\ref{fig:level_agent}, \textit{Fully agentic LLM}). When such agentic LLMs are chained or combined to more complex systems, they add ``learning'' from other agents' output as an essential feature of societal, dynamic behavior. Interacting and observing other agents can thus influence the agent's behavior and, especially for \textit{Complex Adaptive Systems}, agents' interaction lead to adaptive changes in their profiles, just like opinions, and individual behavior changes due to other social actors' opinions or behavior, mimicking fully dynamic social interactions \citep{parkEnhancingAnomalyDetection2024}. The difference between Level 4 and 5---as not visible in Figure~\ref{fig:level_agent}---is the complexity of agentic systems on Level 5, which leads to \textit{emergent} behavior based on the agents' interactions.

This spectrum is not merely theoretical. A concrete example can help illustrate the functional distinctions: In an email triage task, a Level 0 system (\textit{LLM-as-Tool}) might classify or summarize individual messages on demand but lacks any contextual awareness or memory across turns. A Level 1 system (\textit{LLM-as-Role}) is prompted to simulate a consistent persona—say, an executive assistant—with stylized preferences and tone, responding in a role-consistent way but without memory beyond the current prompt. A Level 2 \textit{Agent-like LLM} might retain state across sessions, learn user preferences over time, and flag or draft replies based on simple prioritization rules. By Level 3, a \textit{Fully agentic LLM} could autonomously monitor the email stream, integrate sender context, resolve calendar conflicts, and initiate follow-ups—without requiring user prompting. At Level 4, multiple such agents could coordinate inbox management for different roles within an organization (e.g., assistant, PR, legal), negotiating task allocation and sharing knowledge. Finally, at Level 5, entire ecosystems of agents could simulate broader organizational communication flows, emergent work norms, or institutional adaptation under varying external pressures.

\subsection{Architectural Components for LLM-based Agents}

Agentic LLM systems typically incorporate a set of architectural components that enable them to operate beyond reactive text generation. These components evolve across the six levels described in Table~\ref{tab:llm_agent_architecture}, supporting increasingly autonomous, interactive, and emergent behavior.

At the lower levels (0--2), systems range from stateless LLM tools to agent-like entities capable of maintaining session memory and executing goal-directed tasks. Here, architectural requirements include prompt engineering, session-based memory buffers, and basic control logic. At Level 3, fully agentic LLMs incorporate persistent memory stores, strategic planning capabilities, environmental interfaces, and tool-use APIs, allowing them to perceive and interact with digital environments. These capabilities support full engagement with the OODA loop---enabling the agent to observe, orient, decide, and act in context-sensitive ways. At Level 4, multi-agent systems require additional infrastructure for inter-agent communication, task division, negotiation, and shared goal alignment. Architectures must support parallelism, role differentiation, and message-passing protocols to coordinate multiple agents within a shared task environment. At Level 5, complex adaptive systems add yet another layer: mechanisms for learning, feedback loops, and emergent behavior. These require architectures capable of tracking population-level patterns, adapting strategies over time, and supporting norm formation or diffusion through decentralized interactions.

The construction of these systems typically follows modular design patterns widely discussed in recent system surveys \citep{wangSurveyLargeLanguage2024,gaoLargeLanguageModels2024}. A central LLM often serves as the cognitive core, orchestrating inputs and outputs across supporting modules; this is not strictly necessary, especially in fully-decentralized setups. Tool-use APIs extend the LLM's reach beyond language (and often vision in multi-modal LLMs), enabling direct manipulation of software environments (e.g., web browsing, data querying, programming, or GUI control). Memory is implemented through vector databases, prompt histories, or structured knowledge graphs, enabling contextual continuity across interactions \citep{zhangSurveyMemoryMechanism2024}. Planning capabilities are scaffolded by frameworks such as ReAct \citep{yaoReActSynergizingReasoning2023}, AutoGPT \citep{firatWhatIfGPT42023}, and LangChain \citep{kok-shunIntertwiningTwoArtificial2023}, which provide structures for multi-step reasoning, goal decomposition, and iterative feedback integration \citep{huangUnderstandingPlanningLLM2024}. At higher levels, orchestration layers coordinate perception, planning, execution, and learning cycles across multiple agents, supporting emergent behavior and adaptive simulation within complex social environments.

As LLM systems evolve along this continuum, they acquire the ability not only to execute predefined tasks but also to generate strategies, evaluate outcomes, and modify behavior in response to environmental changes. This is particularly evident at higher levels of the continuum---Levels 4 and 5---where multiple agentic entities operate in coordinated fashion. In such multi-agent configurations, coordination mechanisms become essential: agents must communicate, negotiate, align goals, and resolve conflicts. According to \cite{wangSurveyLargeLanguage2024}, these coordination mechanisms can take various structural forms, ranging from centralized planning (via a root controller) to fully decentralized interaction paradigms where agents dynamically exchange messages, assign subtasks, or vote on decisions. Some systems even adopt federated structures, mirroring principles of institutional governance and self-organization. Thus, these multi-agent configurations mirror established principles in distributed cognition, where knowledge and problem-solving are not confined to an individual but are distributed across multiple interacting agents and artifacts \citep{zhangDistributedCognitionRepresentation2006}. Moreover, the dynamics observed in these systems can and have been studied via complexity theory and studies of collective intelligence, which emphasize emergence, feedback loops, and adaptive restructuring \citep{epsteinGenerativeSocialScience2012, duenez-guzmanSocialPathHumanlike2023}. Architecturally and behaviorally, LLM-based agentic systems are increasingly capable of supporting simulations that model not only individual reasoning but also organizational behavior, norm evolution, and systemic adaptation. 

By clarifying these developmental stages and their architectural dependencies---each aligned with OODA loop phases---the framework that we propose contributes to a more nuanced understanding of agentic potential in LLM-based systems. These distinctions are foundational not only for guiding technical implementation but also for shaping how researchers theorize, validate, and ethically engage with increasingly autonomous computational agents in the social sciences.

\section{Relevance for Social Sciences}

Recent scholarship has begun to systematically explore the implications of LLMs for the social sciences, both as methodological tools and as objects of inquiry. A foundational entry point is provided by \cite{valdenegroLLMDigestSocial2023}, who demystifies LLM architectures for social scientists and emphasizes the importance of epistemological caution. LLMs, he argues, are not reasoning agents but pattern-completion engines, trained on extensive textual corpora to predict plausible next tokens. This framing positions LLMs as probabilistic simulators rather than interpretable decision-makers, underscoring the need for methodological transparency and restraint in their use. However, \cite{valdenegroLLMDigestSocial2023} also highlights the practical affordances of LLMs in the social sciences: they enable low-cost data augmentation, scalable qualitative analysis, and novel forms of simulation-based inquiry. This dual potential, as both powerful analytical tools and ethically complex objects of study, has accelerated the integration of LLMs into social science research.

\begin{longtable}{p{0.5cm}p{1.3cm}p{5.4cm}p{7cm}}
\toprule
\textbf{Lvl} & \textbf{System Type} & \textbf{Social Science Focus} & \textbf{Scientific Sources} \\
\midrule

\multirow{4}{*}{0} & \multirow{4}{=}{LLM-as-Tool} 
& Text, idea, material generation & \cite{gaoLargeLanguageModels2024, keExploringFrontiersLLMs2024,venkatramanCollabStoryMultiLLMCollaborative2025} \\
& & Qualitative coding & \cite{ziemsCanLargeLanguage2024,nilssonAutomaticImplicitMotive2025, keExploringFrontiersLLMs2024, demszkyUsingLargeLanguage2023} \\

& & Data analysis & \cite{organisciakSemanticDistanceAutomated2023, keExploringFrontiersLLMs2024,haaseSDATMultilingualGenAIDriven2025} \\
& & Literature review & \cite{karjusMachineassistedQuantitizingDesigns2025, keExploringFrontiersLLMs2024} \\
\midrule

\multirow{3}{*}{1} & \multirow{3}{=}{LLM-as-Role}
& Persona simulation (Big Five, etc.) & \cite{wangEvaluatingAbilityLarge2025,huangDesigningLLMAgentsPersonalities2024,liQuantifyingAIPsychology2024,sunRandomSiliconSampling2024} \\
& & Emotional behavior simulation & \cite{mozikovGoodBadHulklike2024} \\
& & Human preferences simulation & \cite{zhangGenerativeAgentsRecommendation2024,santurkarWhoseOpinionsLanguage2023,rossiProblemsLLMgeneratedData2024} \\
\midrule

\multirow{3}{*}{2} & \multirow{3}{=}{Agent-like LLM}
& Experiment replication & \cite{filippasLargeLanguageModels2024,hortonLargeLanguageModels2023,yeykelisUsingLargeLanguage2024,hewittPredictingResultsSocial2024,lippertCanLargeLanguage2024,mozikovGoodBadHulklike2024,zhangGenerativeAgentsRecommendation2024} \\
& & Survey response simulation & \cite{aherUsingLargeLanguage2023,argyleOutOneMany2023,tjuatjaLLMsExhibitHumanlike2024,vonderheydeAssessingBiasLLMGenerated2023} \\
& & Measurement and assessment & \cite{wangCanLLMsReplace2025,brickmanLargeLanguageModels2025} \\

\midrule
\multirow{2}{*}{3} & \multirow{2}{=}{LLM-based Agents}
& Problem solving & \cite{zhangCreativeAgentsEmpowering2023,liuTwoHeadsAre2024} \\
& & Autonomous experimentation & \cite{boikoEmergentAutonomousScientific2023,manningAutomatedSocialScience2024,staracePaperBenchEvaluatingAIs2025,yamadaAIScientistv2WorkshopLevel2025} \\
\midrule

\multirow{6}{*}{4} & \multirow{6}{=}{Multi-Agent Systems}
& Research team & \cite{sankaranarayananAutomatingThematicAnalysis2025, gottweisAcceleratingScientificBreakthroughs2025, gottweisAICoscientist2025} \\
& & Scientific exploration & \cite{karjusMachineassistedQuantitizingDesigns2025,raskAverageExploringPotential2024} \\
& & Debating teams & \cite{estornellMultiLLMDebateFramework2024, flaminoLimitsLargeLanguage2025} \\
& & Collaborative task-solving & \cite{liMetaAgentsSimulatingInteractions2023,phelpsMachinePsychologyCooperation2023, zhangExploringCollaborationMechanisms2024, fengWhenOneLLM2025, liCAMELCommunicativeAgents2023, xuExploringLargeLanguage2024} \\
& & Psychological assessment & \cite{yangPsychoGATNovelPsychological2024, kjellRatingScalesTargeted2024} \\
& & Learning communities & \cite{chuLLMAgentsEducation2025, soltoggioCollectiveAILifelong2024, zhangSimulatingClassroomEducation2024,zhangEduPlannerLLMBasedMultiagent2025} \\
\midrule

\multirow{5}{*}{5} & \multirow{5}{=}{Complex Adaptive System}
& Human-like social network behavior & \cite{parkGenerativeAgentsInteractive2023, gaoS3SocialnetworkSimulation2023, luLLMsGenerativeAgentbased2024} \\
& & Emergent social dynamics & \cite{lengLLMAgentsExhibit2024, wuLLMBasedEmpatheticResponse2024, chenAgentverseFacilitatingMultiagent2023, daiArtificialLeviathanExploring2024, yuanMeasuringSocialNorms2024, wangInvestigatingExtendingHomans2025, demszkyUsingLargeLanguage2023} \\
& & Collaboration and competitive dynamics & \cite{chenAgentverseFacilitatingMultiagent2023, chanScalableEvaluationCooperativeness2023, piattiCooperateCollapseEmergence2024,zhaoCompeteAIUnderstandingCompetition2024} \\
& & Opinion dynamics & \cite{chuangSimulatingOpinionDynamics2024, liPoliticalLLMLargeLanguage2024, cisneros-velardePrinciplesOpinionDynamics2024, piaoAgentSocietyLargeScaleSimulation2025} \\
& & Psychological and health dynamics & \cite{kambeitzModellingImpactEnvironmental2025, wuLLMBasedEmpatheticResponse2024} \\
\bottomrule
\caption{Examples of LLM-Based Systems Across Agentic Levels in Social Science Research}
\label{tab:llm_agentic_examples}

\end{longtable}

In parallel, scholars like \cite{filippasLargeLanguageModels2024} and \cite{aherUsingLargeLanguage2023} emphasize that LLMs can serve as proxies for human reasoning in computational experiments. They argue that LLMs, despite their non-human cognitive architecture, capture many aspects of human reasoning due to their extensive training on vast, diverse text corpora. This implicit computational model of human language and behavior allows LLMs to approximate social cognition, making them powerful stand-ins for human participants in certain experimental contexts. As \cite{rossiProblemsLLMgeneratedData2024} notes, ``since LLMs are trained on massive amounts of online data, the data will be able to capture fine details of the social system and of the several populations in it'' (p. 153), positioning them as potentially invaluable tools for studying human social dynamics at scale.

The rapid evolution of LLM-based agentic systems---ranging from stateless text generators (Level 0) to fully autonomous, self-organizing systems (Level 5)---has already begun to reshape how social phenomena can be modeled, simulated, and interpreted. These systems are not merely replacing traditional computational methods but are enabling qualitatively new forms of socio-technical modeling. By embedding memory, task-specific reasoning, and adaptive behavior into LLMs, researchers have created multi-agent systems capable of maintaining state, acting autonomously, coordinating with other agents, and interacting with digital environments \citep{wangSurveyLargeLanguage2024}. This progression reflects a fundamental shift in how social phenomena can be studied: from isolated, context-free text processing to dynamic, context-rich simulations of human behavior and interaction.

Crucially, the examples provided in this paper, summarized in Table~\ref{tab:llm_agentic_examples}, are not intended as an exhaustive or definitive list. Instead, they illustrate the underlying principles of each agentic tier, serving as both proof-of-concept demonstrations and inspiration for further empirical work. As the field evolves, we anticipate that many more examples will emerge, reflecting the ongoing innovation in LLM-based social modeling. To provide a clearer understanding of this progression, the following subsections explicitly describe each tier, moving from the foundational LLM-as-Tool systems, through increasingly complex configurations that incorporate memory, autonomy, and multi-agent interactions, up to fully adaptive, self-organizing systems capable of simulating societal dynamics. These examples not only validate the proposed framework but also reveal the accelerating relevance of LLM agents for the design and interpretation of computational social science.

\subsection{Level 0: LLM-as-Tool}

At the foundational level, LLMs serve as stateless tools for generating, summarizing, and transforming text. While they do not exhibit agentic properties such as memory, autonomy, or environmental awareness, their capacity to produce high-quality, contextually appropriate language makes them valuable instruments in social science workflows. Most directly, they function as productivity aids: researchers increasingly rely on LLMs to generate survey items, synthesize literature, and create stimulus materials for experiments \citep{keExploringFrontiersLLMs2024, demszkyUsingLargeLanguage2023}. Their ability to generate plausible, well-structured text at scale supports not only experimental design but also the development of instructional or exploratory materials in educational and research settings \citep{thapaLargeLanguageModels2025}. Beyond automation, LLMs are also adopted as cognitive partners, tools for ideation and structured brainstorming. Researchers have begun to use LLMs in early-stage project scoping, conceptual modeling, and creative hypothesis development. These ``thinking companions'' can provide useful analogies, suggest alternative framings, or assist in exploring conceptual distinctions \citep{boersExploringCognitiveStrategies2025}. Even without memory or reasoning capabilities, their probabilistic synthesis of prior language data makes them powerful tools for lateral thinking and divergent idea generation \citep{haaseArtificialMusesGenerative2023, haaseHasCreativityLargeLanguage2025}. As an example of text generation, the CollabStory framework demonstrates how multiple LLMs can be chained together to co-author fictional narratives collaboratively. This approach reveals critical methodological challenges for authorship attribution and narrative coherence, as the combined outputs of distinct LLMs can exhibit complex, hard-to-disentangle writing patterns. While not truly agentic, these systems stretch the boundaries of what constitutes authorship and collaborative creativity in AI-generated content \citep{venkatramanCollabStoryMultiLLMCollaborative2025}.

Another area of impact lies in qualitative coding and text classification. Zero-shot prompting of LLMs has been shown to produce taxonomic labeling and free-form explanations that rival, and occasionally surpass, those of human coders in terms of clarity and interpretability \citep{ziemsCanLargeLanguage2024}. For instance, implicit motive coding, traditionally labor-intensive, can be reliably automated with LLMs, achieving accuracy on par with expert annotators while reducing processing time by over 99\% \citep{nilssonAutomaticImplicitMotive2025}. These advances position LLMs as scalable alternatives for coding open-ended responses or applying theory-based classifications to large textual datasets. LLMs are also becoming central to literature review processes. They help researchers extract, summarize, and organize insights from expansive textual corpora, a task traditionally requiring significant manual effort. In mixed-methods designs, LLMs are increasingly used to support ``quantitizing'' workflows---that is, converting qualitative insights into structured forms suitable for statistical analysis \citep{karjusMachineassistedQuantitizingDesigns2025,organisciakSemanticDistanceAutomated2023}. In psychology and related disciplines, LLMs now assist in hypothesis generation, experimental design, and methodological instruction \citep{keExploringFrontiersLLMs2024, demszkyUsingLargeLanguage2023}.  While LLMs are best understood as probabilistic text generators rather than reasoning agents, their integration into research pipelines marks a pivotal shift. They augment the researcher's ideational capacity, reduce manual load, and introduce new possibilities for how inquiry is initiated and iterated in social science. This level, though minimal in agentic complexity, lays the groundwork for more interactive and autonomous systems examined in higher levels of the continuum.

\subsection{Level 1: LLMs as Role-Taker}

At Level 1 of the agentic continuum, LLMs transition from general-purpose tools to entities capable of maintaining pre-defined roles or personas. These systems remain stateless and externally controlled, but are prompted to exhibit consistent behavioral or psychological patterns, such as personality traits, preferences, or affective dispositions. This involves ``conditioning'' the model through carefully structured prompts or system messages, often supplemented by role descriptions, trait parameters, or simulated contexts. Unlike Level 0 tools, these LLMs are not just language generators; they are simulations of particular kinds of people or profiles, meant to behave coherently across tasks in line with their assigned characteristics. For the social sciences, this opens new opportunities for scalable, repeatable simulations of human-like behavior under controlled conditions. LLMs can serve as synthetic participants in experiments that examine individual differences, group dynamics, or the effects of psychological traits on decision-making. Researchers can explore how simulated agents with distinct personalities respond to the same scenarios, or how demographic conditioning (e.g., gender, political affiliation, or cultural background) influences language-based responses. These persona-driven applications allow for finer control in experimental setups and offer cost-effective ways to model individual-level variability, especially when studying sensitive or ethically difficult issues.

Concrete use cases already demonstrate this potential. A growing body of work focuses on simulating stable personality traits, like the Big Five dimensions. LLMs have been prompted to adopt personality profiles and exhibit trait-consistent behavior with high internal coherence and convergent validity compared to human self-report data \citep{huangDesigningLLMAgentsPersonalities2024, wangEvaluatingAbilityLarge2025}. Similar approaches explore how emotion simulation can influence decision-making: prompting an LLM with affective framing alters its behavioral outputs in line with human emotional responses \citep{mozikovGoodBadHulklike2024}. Other studies extend role simulation to psychometric modeling. LLMs have been evaluated for their ability to express coherent patterns across psychological dimensions like motivation, affect, and decision-making tendencies, showing potential for systematic, repeatable behavioral outputs in simulated settings \citep{liQuantifyingAIPsychology2024}. This opens the door for using LLMs in tasks traditionally reserved for human participants, such as theory testing or applying interventions in experimental psychology and sociology.

In applied domains, LLMs are increasingly used to simulate user profiles and decision-making behavior. For example, persona-based agents have been designed to mimic MovieLens or Amazon-Book users, incorporating memory and emotional modeling to evaluate how personalized recommendations are interpreted or filtered \citep{zhangGenerativeAgentsRecommendation2024}. These user simulators provide insight into the dynamics of algorithmic influence, filter bubbles, and preference formation. However, caution is warranted. Some findings show that even when LLMs are prompted to represent distinct demographic or political groups, the resulting responses often reflect generic or homogenized viewpoints. This is partially attributed to the \textit{Reinforcement Learning from Human Feedback} (RLHF) process, which tends to smooth out controversial or sensitive positions, thus leading to reduced representational fidelity for underrepresented or marginalized groups \citep{santurkarWhoseOpinionsLanguage2023, rossiProblemsLLMgeneratedData2024}. Although demographic conditioning (e.g., ``random silicon sampling'') can generate subgroup-like distributions \citep{sunRandomSiliconSampling2024}, there remains an epistemic and ethical risk in treating these simulations as proxies for real-world populations.

In sum, Level 1 systems simulate role-specific behavior without autonomous goal selection or memory. Their primary affordance lies in modeling consistent, human-like responses for controlled experimental or design settings—especially in psychology, marketing research, and human-AI interaction studies.

\subsection{Level 2: Agent-like LLM}

At Level 2, LLMs move beyond simple role simulation toward more autonomous, structured task performance. These systems exhibit ``agent-like behavior''; thus they can decompose tasks, make context-sensitive decisions, and access memory or external tools when necessary. While they lack full autonomy or environment interactivity, they are increasingly used as self-contained agents capable of acting purposefully within constrained scenarios. Architecturally, these LLMs may operate with internal memory buffers, tool integration, and iterative reasoning chains (e.g., chain-of-thought or self-reflection prompts). They can also exhibit planning behaviors, simulate deliberation, and interact with APIs or local documents when embedded in systems with expanded action spaces \citep{wangSurveyLargeLanguage2024}.

In the social sciences, these agent-like systems serve a vital function: they can emulate human decision-making and support scalable experimentation without requiring continuous oversight. Instead of merely responding statically, they begin to act as semi-autonomous evaluators, actively interpreting, quantifying, and shaping human-like constructs. Their applications span simulation of behavioral experiments, synthetic survey responses, and even the generation or evaluation of psychological or sociological constructs.

\subsubsection*{Experiment Replication}

One prominent application of agent-like LLMs is the replication of classic experiments. Researchers have shown that these systems can simulate behavioral game-theoretic interactions by adopting specific preferences and strategies, replicating known experimental effects such as altruism or fairness \citep{filippasLargeLanguageModels2024, hortonLargeLanguageModels2023}. In marketing research, LLM agents have replicated over 130 media effect studies with a high degree of correspondence to human results, offering a scalable pathway to validate empirical findings and test generalizability \citep{yeykelisUsingLargeLanguage2024}. Likewise, LLMs have demonstrated strong forecasting capabilities, matching or exceeding expert predictions in social and behavioral science domains \citep{hewittPredictingResultsSocial2024, lippertCanLargeLanguage2024}.

These use cases underscore a key epistemic shift: LLMs are not just tools for supporting human analysis but can serve as experimental subjects or forecasters in their own right, enabling low-cost replication and hypothesis testing. Still, caution remains essential; while models may match behavioral outcomes on average, they often miss interaction effects or contextual nuances that humans intuitively grasp.

\subsubsection*{Survey Response Simulation}

LLMs are also used to simulate human survey responses across various domains—from consumer behavior to political science \citep{aherUsingLargeLanguage2023, argyleOutOneMany2023}. Through demographic conditioning or persona steering (cf. Level 1), they can mimic subgroup-specific attitudes and predict, for example, voting behavior \citep{vonderheydeAssessingBiasLLMGenerated2023}. Yet, while these approaches show promise, limitations persist: LLMs trained with RLHF tend to produce overly sanitized, bias-suppressed outputs, leading to underrepresentation of real-world variance in attitudes or social biases \citep{santurkarWhoseOpinionsLanguage2023, tjuatjaLLMsExhibitHumanlike2024}. This calls into question their validity as stand-ins for marginalized or complex social groups \citep{rossiProblemsLLMgeneratedData2024}.

Other studies focus on building psychologically plausible agents, combining LLMs with stance detection and cognitive architectures to simulate adaptive human behavior within agent-based models, aiming to overcome the abstraction limitations of traditional agent-based models \citep{mitsopoulosPsychologicallyValidGenerativeAgents2023}. For example, \citet{yangLLMMeasureGeneratingValid2024} introduce \textit{LLM-Measure}, a prompting framework that enables LLMs to generate high-quality survey items aligned with psychometric standards, demonstrating that LLM-generated items can match or exceed human-written ones in validity, consistency, and representativeness across diverse psychological constructs. Additionally, LLMs have been used to generate ``silicon samples'' for early-stage research, offering realistic but ethically unencumbered approximations of consumer behavior \citep{argyleOutOneMany2023,sarstedtUsingLargeLanguage2024}.

\subsubsection*{Measurement and Assessment}

Beyond simulating human input, agent-like LLMs are increasingly used to assess and quantify psychological and sociological constructs. Recent work demonstrates how LLMs can serve as embedded evaluators within structured research pipelines: for instance, LLMs have been assigned the roles of coders and analysts in experimental setups to interpret qualitative inputs, simulate inter-rater discussion, and generate hypothesis-driven insights—mirroring key steps of empirical inquiry in psychology and the social sciences \citep{brickmanLargeLanguageModels2025}. These agentic roles extend beyond static annotation by enabling dynamic role-taking, turn-based deliberation, and methodologically grounded prompt chaining.

A related approach is the emerging field of ``LLM-as-a-judge'', which explores whether LLMs can replace human evaluators in specialized contexts like software engineering. For example, recent empirical studies have evaluated the reliability of LLM-based judgment systems for code review and quality assessment. These systems have demonstrated near-human performance in assessing software artifacts, achieving strong alignment with expert human scores while reducing the need for costly manual evaluation \citep{wangCanLLMsReplace2025}. These findings suggest that LLM-based evaluators can approximate human judgment in complex, high-stakes evaluation tasks, providing a foundation for more autonomous, agent-like assessment systems.

\subsection{Level 3: LLM-based Agents}

At this level, LLMs transition from static tools or role-players into fully operational agents capable of planning, memory retention, and interaction with external systems or environments. These agents no longer respond in isolation; they observe, orient, decide, and act (OODA, \citealt{boydDiscourseWinningLosing2018,osingaScienceStrategyWar2007}) in ways that echo foundational agent-based modeling principles. Classical agent-based models aimed to capture how micro-level behaviors could generate emergent macro-level patterns, thus often abstracting cognition or decision-making rules \citep{epsteinAgentbasedComputationalModels1999, epsteinGenerativeSocialScience2012}. In contrast, LLM-powered agents now make it possible to embed sophisticated language-based reasoning, self-reflection, and interactive planning directly within simulated entities \citep{parkGenerativeAgentsInteractive2023, linAgentSimsOpenSourceSandbox2023}.

Such agents typically incorporate memory streams, persona consistency, planning modules, and action spaces (e.g., tool use, API calls, web access), enabling meaningful adaptation over time. These capabilities bring LLM-based simulations closer to modeling the dynamic, feedback-driven nature of social environments. However, current implementations also face significant challenges. While they excel in controlled, omniscient simulations, they often struggle with uncertainty and asymmetric information---a hallmark of real-world social interaction \citep{zhouThisRealLife2024}. Moreover, their adaptability in distributed networks remains limited compared to human actors, particularly in environments requiring cooperation or trust \citep{hanStaticNetworkStructure2024}. Nevertheless, the emerging field of LLM-based agent simulations shows immense promise. It is rapidly expanding across domains such as computational social science, digital experimentation, and complex system design \citep{gaoLargeLanguageModels2024}.

Finally, it is important to note that this level marks a conceptual turning point in the continuum: agents become capable of autonomous planning and social reasoning, but typically operate in relative isolation or only loosely coordinated teams. The following subsections illustrate key use cases in social science research.

\subsubsection*{Problem Solving Agents}

LLM-based agents are increasingly designed for creative and autonomous problem-solving in both text-based and embodied environments. One prominent example involves agents operating within a Minecraft simulation, where an imagination module, powered by an LLM, generates multiple candidate responses before acting \citep{zhangCreativeAgentsEmpowering2023}. Given abstract prompts like ``build a bridge'', these agents evaluate diverse structural ideas internally before committing to action. Their performance surpassed baseline agents by demonstrating flexible, goal-aligned creativity that was not hardcoded but emergent from internal planning processes \citep{zhangCreativeAgentsEmpowering2023}. This work introduced a general architecture for embedding imagination in agents and offered new ways of evaluating creativity in open-ended tasks, including novel metrics based on GPT-4.

Another strategy relies on enhancing reasoning through collaboration across multiple LLMs. Instead of using a single model in isolation, multiple prompts or even multiple models are combined sequentially to overcome individual limitations in logic or knowledge coverage. This technique, while not yet a full agentic system, reflects the growing trend of modularizing problem-solving across distributed reasoning paths, suggesting new architectures for collective cognition \citep{liuTwoHeadsAre2024}.

\subsubsection*{Autonomous Experimentation}

Beyond creative tasks, agentic LLMs are now entering domains traditionally reserved for human researchers. In high-stakes scientific experimentation, LLM-based agents have been developed with planning, memory, and access to scientific tools or external data sources. The following examples are not proper cases for social science, however, as other research fields seem more advanced in incorporating agents for research, we post these as a blueprint or inspiration for more social adaptations. One example presents a chemistry-focused agent capable of autonomously designing and evaluating experimental protocols, drawing on external tools via natural language interfaces. Here, LLMs are not just embedded within simulations, but orchestrate complex workflows in real time \citep{boikoEmergentAutonomousScientific2023}. Another, particularly illustrative, example is provided by the \textit{PaperBench} framework, which simulates a team of agents, each assuming a distinct scientific role (e.g., PhD student, reviewer, PI), to autonomously reproduce recent machine learning papers using publicly available resources \citep{staracePaperBenchEvaluatingAIs2025}. While its domain is technical, the system exemplifies how coordinated agentic reasoning, role-based task allocation, and iterative planning can be used to assess scientific validity. In a broader sense, PaperBench demonstrates how agentic LLM systems can serve not only as individual experimenters but as autonomous evaluators and replicators of complex research processes, offering an important conceptual and methodological reference for future applications in computational social science.

A striking example of agentic performance beyond typical task automation comes from DeepMind's AlphaGeometry system, which autonomously solved geometry problems from the \textit{International Mathematical Olympiad} (IMO) at a silver medal level \citep{alphaproofandalphageometryteamsAIAchievesSilvermedal2024}. Combining a neural language model with symbolic deduction capabilities, AlphaGeometry reasoned through abstract, multi-step proofs and did so in a way that matched human expert performance. This underscores the potential of hybrid agentic systems to engage in genuine epistemic innovation across domains---including those, like mathematics, long considered the pinnacle of human reasoning. It also highlights the emerging ability of LLM-based agents to autonomously contribute to knowledge generation in highly structured, logic-driven problem spaces.

In the realm of social science, agentic LLMs are being tasked with hypothesis testing and causal inference. By embedding structural causal models into LLM-based agents, researchers have demonstrated the feasibility of automated theory-building and simulation-based testing of social mechanisms. These systems can generate hypotheses, simulate plausible social scenarios, and interpret results—all without direct human instruction at every step \citep{manningAutomatedSocialScience2024}. Such capabilities move LLMs beyond mere assistance roles into positions of epistemic contribution, expanding the potential for theory-driven inquiry through generative computation.

\subsection{Level 4: Multi-Agent Systems}

At this stage of development, LLMs are no longer acting alone. Instead, they are embedded into multi-agent systems, where multiple LLM-based agents interact, communicate, negotiate, and coordinate with one another, either as equals or through hierarchical structures. These agents often hold differentiated roles, personas, or goals, and collectively simulate dynamic group behavior. The architectural shift toward agent societies marks a critical leap: it allows for the modeling of collaborative problem solving, scientific co-creation, emergent deliberation, and distributed decision-making---phenomena at the heart of social complexity.

This level brings the coordination logic of agentic AI into full view. From mimicking human research teams and simulating debates to playing communication games or constructing fictional narratives, multi-agent LLM systems unlock entirely new modes of generative social simulation. However, this level also introduces distinct methodological and epistemic risks: agents may converge prematurely due to shared training biases, amplify misinformation, or fail to exhibit genuine behavioral diversity \citep{estornellMultiLLMDebateFramework2024, flaminoLimitsLargeLanguage2025}. Still, the promise is profound: Level 4 systems provide a living laboratory for studying social dynamics---scalable, repeatable, and manipulable in ways human-only teams cannot match.

\subsubsection*{Research Team}

Multi-agent systems are increasingly being deployed to emulate the collaborative workflows of human researchers. For example, the multi-agent system proposed by \cite{sankaranarayananAutomatingThematicAnalysis2025} automates thematic analysis by orchestrating multiple LLM sub-agents to mirror the traditionally manual, iterative steps of qualitative data coding. This approach allows for autonomous data preprocessing, codebook development, consensus building, and final theme generation, significantly enhancing the transparency and scalability of TA processes. The system also incorporates mechanisms for managing ambiguity and ensuring consistency across agent outputs, aligning closely with human expert evaluations in initial benchmarks. Similarly, Google's Gemini 2.0 powers the ``AI Co-Scientist'', a system capable of autonomously formulating, testing, and refining hypotheses in biomedical domains. This system illustrates the viability of LLM-based multi-agent architectures for dynamic scientific discovery, demonstrating that agentic LLM systems can move beyond static task execution to adaptive, hypothesis-driven inquiry \citep{gottweisAcceleratingScientificBreakthroughs2025, gottweisAICoscientist2025}.

A further development of this concept is embodied by Sakana AI's \textit{AI Scientist} \citep{luAIScientistFully2024}, a multi-agent framework designed to emulate the division of labor, collaboration, and role diversity within real-world research teams. The system orchestrates heterogeneous agents—ranging from hypothesis generators and experimental designers to result analysts and synthesis agents—across iterative research loops. Notably, the AI Scientist incorporates principles of diversity and redundancy, enabling agents with different architectures and learning biases to challenge and refine one another's outputs. This design mirrors the epistemic pluralism often found in productive human teams and leads to improved robustness and creativity in scientific problem-solving. By coordinating diverse LLM-based agents in structured yet flexible workflows, Sakana's system demonstrates how agentic teams can replicate and even enhance human-style collective intelligence in research settings.

\subsubsection*{Scientific Exploration}
LLM-based multi-agent systems are also being used to explore scientific discovery processes themselves. These systems can simulate full research pipelines---from literature review to hypothesis generation, experimental design, and evaluation---offering a new paradigm for automating or augmenting scientific workflows. One line of work proposes a systematic framework for integrating LLMs into qualitative research processes, enabling mixed-methods analysis and cross-linguistic hypothesis exploration. These systems function less as autonomous researchers and more as collaborative agents augmenting human capabilities, reinforcing the view that LLMs should complement, not replace, domain experts \citep{karjusMachineassistedQuantitizingDesigns2025}. In a broader case study of LLM deployment across the entire research cycle, another study examines real-world practices and institutional constraints when integrating LLMs into social science workflows. The findings emphasize both the epistemic affordances and governance challenges of using generative models for scientific knowledge production \citep{raskAverageExploringPotential2024}.

\subsubsection*{Debating Team}

LLM-based agents are increasingly used in simulated deliberative contexts, offering new tools for studying group dynamics, argumentation, and decision-making. Studies show that when tasked with debating issues or reaching consensus, agents can emulate classic social phenomena such as conformity, polarization, and groupthink. For instance, \citet{estornellMultiLLMDebateFramework2024} demonstrate that even in controlled debate simulations, agents develop distinct conversational strategies and coordinate in ways that mirror human group behavior. However, they may also converge on biased outputs due to homogeneous training data and architectural similarities, which can limit argumentative diversity.

A recent experimental study by \citet{flaminoLimitsLargeLanguage2025} examined how LLM agents perform in mixed human-AI debates. In a structured opinion consensus game, agents powered by GPT-4 and Llama 2 interacted anonymously with human participants to discuss topics like climate-conscious diets. The agents stayed consistently on-topic, increased deliberative structure, and helped improve the overall productivity of discussions. Yet, they were less persuasive than human peers---humans were six times more likely to influence one another's opinions. Agents also changed their own views more frequently, suggesting a flexible but less assertive engagement style. While rated as less confident and convincing, their contributions were goal-oriented and constructive, supporting the deliberative process. These findings highlight both the promise and current limitations of LLM agents in modeling social interaction.

\subsubsection*{Collaborative Task Solving}

In more applied contexts, LLM agents are being used to simulate collaborative task environments where social coordination, negotiation, and goal-aligned interaction are critical. For instance, the MetaAgents framework introduces a multi-agent setup that evaluates coordination strategies in job fair simulations, demonstrating that agent groups can autonomously assign roles, negotiate outcomes, and simulate realistic professional interactions \citep{liMetaAgentsSimulatingInteractions2023}. Moreover, recent work has shown that agent differentiation by personality traits and cognitive styles can replicate human-like collaboration dynamics in domains like chess, debate, and collective decision-making. These multi-agent societies display emergent behaviors such as conformity, leadership, and conflict resolution, closely aligning with foundational social psychology concepts \citep{zhangExploringCollaborationMechanisms2024}.

To enable such complex interactions, advanced prompting techniques like inception prompting have been introduced, which embed structured roles and shared goals into the agents' context to facilitate coherent, multi-turn collaboration. This approach has been applied to complex communication games like Werewolf, where agents must infer the intentions of others, develop trust, and navigate high-stakes social interactions without parameter fine-tuning \citep{xuExploringLargeLanguage2024}. Further, the \textit{CAMEL} framework by \cite{liCAMELCommunicativeAgents2023} uses inception prompting to facilitate autonomous cooperation in multi-agent systems. This approach relies on predefined role assignments, structured task specification, and iterative message passing to ensure agents stay on task and avoid role flipping or conversational loops, significantly enhancing task completion efficiency. Such frameworks provide valuable testbeds for studying group dynamics, decision-making, and leadership emergence in synthetic social environments \citep{fengWhenOneLLM2025}.

\subsubsection*{Psychological Assessment}

Multi-agent systems are increasingly being applied to psychological measurement contexts, and recent work has emphasized the potential of LLMs to move beyond conventional rating scales entirely. Rather than relying on fixed-response formats, these systems aim to quantify psychological constructs directly from free-form, natural language. This shift addresses a critical limitation of traditional assessments: the reduction of complex psychological states to a fixed set of numerical ratings. For example, open-ended language responses capture more nuanced, context-rich information about mental states, significantly outperforming conventional rating scales in measures like self-information and context sensitivity. LLMs can achieve near-theoretical upper limits of accuracy in aligning with human-rated scales, while also providing deeper insights into psychological constructs through dynamic, personalized language interpretation \citep{kjellRatingScalesTargeted2024}. Further, the PsychoGAT framework leverages LLM agents to transform traditional self-report scales into interactive fiction games, providing a more engaging and personalized assessment experience for constructs like depression, personality, and cognitive distortions. By integrating role-playing, memory, and dynamic dialogue, this approach achieves strong psychometric validity while enhancing user engagement through game mechanics, improving participant satisfaction and immersion \citep{yangLLMMeasureGeneratingValid2024}. These approaches open new pathways for assessing mental health in more ecologically valid, context-sensitive ways.

\subsubsection*{Learning Communities and Education}

Finally, multi-agent systems are increasingly explored as collective intelligence frameworks for educational contexts. These systems envision networks of LLM agents that can learn from experience, exchange information across nodes, and collectively build a shared knowledge base. This approach aligns with foundational concepts in complexity science, which emphasize decentralized learning, self-organization, and emergent behavior \citep{soltoggioCollectiveAILifelong2024}. Individual LLM-based agent systems show their ability to take on specific roles in the overall learning process (e.g., like an error-detector, \citealt{xuAIDrivenVirtualTeacher2025}, learning-content creator, \citealt{elkinsHowTeachersCan2024}, and automated answer scoring, \citealt{liAutomatedExplainableEducational2025}). \cite{chuLLMAgentsEducation2025} extend this perspective by examining the potential of multi-agent systems for automated tutoring and adaptive instruction. It highlights how LLM agents can collaboratively simulate classroom interactions, integrate feedback loops, and perform real-time performance assessments, creating adaptive, student-centered learning environments. 

As a concrete example, the \textit{SimClass} framework demonstrates how multi-agent LLM systems can replicate classroom dynamics, enabling real-time adaptation to student needs and facilitating personalized instruction through role-based agents \citep{zhangSimulatingClassroomEducation2024}. This approach leverages multi-agent architectures to model the varied cognitive and emotional states of students, offering a scalable platform for studying educational processes at both individual and group levels. In a similar vein, the \textit{EduPlanner} system uses a multi-agent approach to automate the entire instructional design cycle, including content generation, evaluation, and optimization \citep{zhangEduPlannerLLMBasedMultiagent2025}. EduPlanner employs a Skill-Tree structure to represent students' prior knowledge and learning progress, dynamically adjusting lesson plans based on real-time performance feedback. This allows for the generation of personalized educational content, integrating multiple agents for instructional assessment, optimization, and error analysis \citep{zhangEduPlannerLLMBasedMultiagent2025}. The system's iterative, adversarial collaboration among evaluator, optimizer, and analyst agents mimics human teacher-student interactions, reinforcing the concept of collective learning within artificial educational ecosystems. Together, these projects highlight the potential of multi-agent LLM systems to revolutionize educational practice, transforming classrooms into adaptive, learner-centered environments that scale with student needs.

\subsection{Level 5: Complex Adaptive Systems}

At Level 5, LLM-based agent systems extend beyond small-scale collaboration or structured task execution to model entire societies, ecosystems, or complex adaptive systems. These platforms aim to capture the emergent, large-scale dynamics of social systems by integrating hundreds or even thousands of interacting agents, each equipped with memory, autonomous decision-making, and social learning capabilities. Unlike the more constrained, role-based interactions of Level 4, Level 5 agents operate within fluid, adaptive networks where emergent phenomena---such as social norms, power structures, and collective behaviors---emerge spontaneously from the decentralized interactions of individual agents. This transition reflects a fundamental shift from deterministic simulation to adaptive, emergent modeling, enabling researchers to study not just predefined rules but the self-organizing principles that govern macro-social phenomena. Here, the focus shifts from explicit, top-down control to the bottom-up processes that shape the formation of norms, collective intelligence, and long-term social adaptation, capturing the rich, often unpredictable dynamics of real-world societies.

\subsubsection*{Human-Like Social Network Behavior}

Simulating human-like social network behavior is a critical step toward understanding complex social dynamics. Unlike isolated, single-agent simulations, social network systems aim to replicate the intricate web of human interactions at both individual and population levels. These systems capture not only the direct actions of agents but also the emergent phenomena that arise from their interactions, such as information diffusion, collective sentiment shifts, and social contagion.

One pioneering approach is the \textit{Generative Agent} platform, which models small-scale, LLM-driven societies within 2D game environments \citep{parkGenerativeAgentsInteractive2023}. Here, agents autonomously plan their daily routines, engage in conversations, and adapt their behaviors based on prior interactions. These agents exhibit complex emergent behaviors like spontaneous information diffusion, cooperative group dynamics, and even basic forms of political organization, such as hosting mayoral elections and organizing social gatherings \citep{parkGenerativeAgentsInteractive2023}. The system demonstrates the potential for LLMs to simulate realistic, context-sensitive social interactions, providing a foundation for more sophisticated social network modeling. Scaling up this concept, \textit{GenSim} introduces a platform for simulating up to 100,000 agents, incorporating error correction and adaptive learning to manage the complexity of massive agent interactions. The system abstracts agent profiles, multi-agent scheduling, and environmental setups to provide a flexible framework for large-scale, realistic social simulations. This approach significantly advances the computational feasibility of modeling large-scale social networks, enabling the study of emergent phenomena at unprecedented scales \citep{tangGenSimGeneralSocial2024}. GenSim supports real-time interaction among thousands of agents, facilitating the study of macro-level social dynamics, from market behavior to political polarization.

Further extending this concept, the \textit{S³} system introduces a framework for simulating social networks with agents that model emotions, attitudes, and interpersonal interactions \citep{gaoS3SocialnetworkSimulation2023}. It captures emergent phenomena such as collective sentiment shifts and social contagion by integrating fine-tuned LLM agents capable of perceiving and responding to their informational environment. The system has been tested with real-world social network data, demonstrating its ability to replicate complex social phenomena like the spread of political attitudes or the evolution of public opinion. This framework highlights the importance of modeling not just the actions of individual agents but also the emergent properties that arise from their collective interactions, providing valuable insights for both theoretical and applied social science research.

\subsubsection*{Emergent Social Dynamics}

Emergent social dynamics in complex adaptive systems capture the spontaneous formation of cooperation, conflict, and collective order. These dynamics arise from the repeated interactions of individual agents, which collectively produce higher-order social structures. The study of such emergent phenomena draws on foundational theories like Social Contract Theory (SCT, \citealt{daiArtificialLeviathanExploring2024}) and Social Exchange Theory (SET, \citealt{wangInvestigatingExtendingHomans2025}), which provide conceptual blueprints for understanding how micro-level exchanges can aggregate into macro-level social order. 

For instance, the \textit{Artificial Leviathan} framework explores the emergence of cooperative norms and governance structures from a ``state of nature'' characterized by conflict and self-interest. In this model, agents engage in strategic decision-making, alliance formation, and resource management, gradually evolving from isolated, competitive behavior to organized, cooperative societies \citep{daiArtificialLeviathanExploring2024}. This approach closely aligns with Hobbes's original vision of the social contract, where rational self-interest drives the formation of the collective order. Similarly, the \textit{SUVA} framework systematically analyzes LLM agents' socially grounded decision-making based on their textual outputs, revealing tendencies toward fairness, reciprocity, and cooperative behavior without explicit priming \citep{lengLLMAgentsExhibit2024}. This approach demonstrates that even without pre-defined social scripts, LLM agents can exhibit prosocial tendencies, aligning their actions based on context rather than rigid, pre-programmed responses.

Homans' SET provides a micro-level foundation for understanding how individual interactions can lead to collective social phenomena. Recent work has adapted SET to LLM-based agents, modeling the six core propositions of SET---success, stimulus, value, deprivation-satiation, aggression-approval, and rationality---within a controlled, multi-agent society. These systems demonstrate the emergence of cooperation, reciprocity, and social norms without explicit hardcoding of behaviors, capturing more nuanced human-like exchanges through cognitive and affective components, such as affinity scores and social value orientations \citep{wangInvestigatingExtendingHomans2025}. Expanding this concept further, the \textit{AgentSociety} framework introduces a large-scale simulation environment for modeling complex, real-world social phenomena. It supports the exploration of topics like polarization, misinformation spread, and economic policy impacts, providing a concrete testbed for social science research. This approach emphasizes scalability and empirical validation, offering a powerful tool for investigating emergent social dynamics in digitally mediated environments \citep{piaoAgentSocietyLargeScaleSimulation2025}.

\subsubsection*{Collaboration and Competitive Dynamics}

Multi-agent systems at this level are not only designed to study cooperative behavior but also to explore the competitive dynamics that emerge in complex social and economic environments. These systems capture the interplay between collaboration and competition, revealing how agents adapt to both supportive and adversarial conditions over time.

One prominent example is the \textit{GOVernance of the Commons SIMulation} (GOVSIM) platform. It is specifically designed to test LLM agents' capacity for sustainable cooperation in common resource dilemmas. Inspired by economic theories of cooperation and the management of shared resources, GOVSIM simulates scenarios like fisheries, pastures, and pollution, where agents must balance short-term gains with long-term sustainability. This approach highlights both the potential and current limitations of LLM-based systems for modeling real-world governance challenges, providing insights into how collective strategies emerge and stabilize in resource-limited environments \citep{piattiCooperateCollapseEmergence2024}.

In parallel, the \textit{CompeteAI} framework introduces a novel approach for modeling competitive dynamics among LLM-based agents. This system simulates a virtual economy where agents, such as restaurant owners, compete to attract customers by dynamically adjusting their strategies based on market conditions and competitor behavior. The agents exhibit emergent market phenomena, including price wars, brand loyalty, and customer segmentation, closely mirroring real-world competitive pressures in economic environments. This framework emphasizes the importance of long-term strategic reasoning and adaptive decision-making, highlighting how local agent interactions can lead to complex, emergent market behaviors at the system level \citep{zhaoCompeteAIUnderstandingCompetition2024}.

\subsubsection*{Opinion Dynamics}

Modeling opinion dynamics is a cornerstone of social science, capturing how collective beliefs, attitudes, and preferences evolve over time within populations. LLM-based agent systems offer powerful tools for simulating these processes, enabling researchers to study phenomena like consensus formation, polarization, and the spread of misinformation in complex social networks.

Recent work by \cite{chuangSimulatingOpinionDynamics2024} introduces a multi-agent framework for simulating opinion dynamics in large-scale social systems. This approach highlights the inherent biases of LLMs towards producing accurate information, which often drives simulated agents toward consensus around scientifically validated positions. However, by introducing specific prompts that activate confirmation bias, the authors were able to reproduce key social phenomena like ideological clustering and opinion fragmentation, providing a more realistic model of human social interactions. Building on this, \cite{liPoliticalLLMLargeLanguage2024} extended the study of opinion dynamics by integrating political polarization mechanisms into LLM-based social simulation platforms. They demonstrated that LLM agents could replicate known polarization effects, such as ideological clustering and opinion reinforcement, when exposed to biased or segmented information environments. Their work underscores the importance of realistic social dynamics, including homophily and network topology, for capturing the full complexity of opinion evolution in digital and physical societies.

Additionally, \cite{cisneros-velardePrinciplesOpinionDynamics2024} explored the underlying principles that shape opinion dynamics within populations of interacting LLMs. They identified several key biases, including a preference for equity-consensus, caution in opinion shifts, and sensitivity to ethical considerations, which collectively influence the distribution of opinions over time. Their findings emphasize the need to account for these biases when designing multi-agent systems for opinion modeling, as they can significantly impact the emergent properties of such systems. Similarly, \cite{piaoAgentSocietyLargeScaleSimulation2025} introduced the \textit{AgentSociety} platform, a large-scale social simulator capable of modeling the complex interplay between social influence, polarization, and collective behavior in populations exceeding 10,000 agents. Here, LLM-based multi-agent systems are being used to simulate the long-term effects of public policies and external shocks by studying the impact of polarization, the spread of inflammatory messages, and the effects of natural disasters on collective behavior. These systems offer a powerful testbed for evaluating policy outcomes, social resilience, and institutional change.

\subsubsection*{Psychological and Health Dynamics}

LLM-based agents are emerging as powerful tools for modeling the complex interplay of psychological and health dynamics, capturing the multifactorial influences of socio-environmental factors on mental health. These systems offer the potential to model interactions at multiple levels—from individual psychological processes to broader community dynamics and societal influences—providing insights that are often ethically challenging to obtain through human experimentation.

For instance, \cite{kambeitzModellingImpactEnvironmental2025} highlight how generative agents can simulate human-like behavior in virtual environments to investigate the effects of environmental and social determinants on mental health. These agents can replicate complex social interactions, model adverse life events, and capture the nuanced effects of urban stressors, such as social deprivation, pollution, and lack of green spaces, on psychological well-being. Importantly, these models can simulate emergent phenomena, such as resilience, which depends on both individual traits and supportive social networks, thus offering a more realistic representation of mental health dynamics than traditional methods. In a similar vein, \cite{wuLLMBasedEmpatheticResponse2024} propose a multi-agent framework specifically designed to integrate diverse psychological theories, including cognitive-behavioral, psychodynamic, and humanistic approaches, to generate more empathetic responses by professionals. Their framework leverages multiple LLMs acting as psychologists, each representing a different therapeutic perspective, and a decision-making agent to select the most contextually appropriate responses. This approach allows for a more nuanced understanding of psychological dynamics by incorporating multiple interpretive frameworks and iterative feedback processes, effectively mirroring real-world therapeutic interactions.

\section{Discussion}

This paper set out to provide a conceptual scaffold for understanding and categorizing LLM-based agents in the social sciences. By introducing a six-tier developmental continuum, we offer a framework that connects technical advancements with core social science interests---from modeling individual cognition to simulating emergent societal phenomena. The goal was not to catalog all existing systems but to distill the structural differences that matter most for empirical, theoretical, and ethical engagement with agentic AI.

Even though many of the studies cited are still in early stages or preprint form, the rapid pace of publication and prototyping indicates a vibrant, self-reflective research landscape. Notably, this paper presents only a sample of current efforts; further examples are already being developed across social science domains. As the boundaries between technical design and social application continue to blur, the need for interdisciplinary research becomes paramount \citep{vladovaWhyWhomHow2024}. Initial use cases already suggest that LLM-based agents could reshape long-standing methodological constraints in the social sciences. They promise a new kind of flexibility---one that enables systematic experimentation at scale, dynamic simulation of collective behavior, and the iterative refinement of theory \textit{in silico}. Rather than simply mimicking human responses, these systems offer opportunities to construct new epistemic instruments: agents that help test, challenge, and expand our understanding of complex social dynamics.

Importantly, while some tiers, especially Levels 0 to 2, are already being widely implemented for automation and efficiency, higher levels that support coordination and emergence remain underexplored but highly promising. Particularly, Level 3 systems appear relatively scarce, likely due to the overhead of maintaining coherent individual agents compared to the greater expressiveness and scalability of multi-agent coordination. Still, as capabilities grow, so too does the opportunity to model not just isolated behaviors but the generative mechanisms of social systems themselves. As this field progresses, it brings with it not only technical innovation but also a strong tradition of critical reflection. Many recent publications have already engaged deeply with issues of bias, validity, and epistemological caution, which is an encouraging sign for the responsible evolution of agentic AI in social science. The following sections take up this balance in more detail, examining both the methodological affordances and the challenges that come with embedding LLM agents into empirical research and simulation.

\subsection{Core Potentials and Methodological Considerations}

The recent proliferation of LLM-based agent research opens unprecedented avenues for the social sciences. These systems offer dual utility: at lower levels (0--2), they serve as scalable instruments for automating routine research tasks; at higher tiers (3--5), they become tools for generating and testing theories of social interaction, coordination, and emergence. At the foundational levels, LLMs enhance efficiency and consistency in established methods. Stateless tools (Level 0) are already widely used for classification, summarization, and synthetic data generation \citep{valdenegroLLMDigestSocial2023}. Role-based systems (Level 1) enable contextualized outputs across longer interactions, supporting more coherent and reproducible workflows \citep{wangSurveyLargeLanguage2024}. Level 2 systems extend this further by incorporating task autonomy and memory, which allow agents to simulate goal-driven behavior and manage multi-step analyses or experimental routines \citep{hewittPredictingResultsSocial2024}.

From Level 3 onward, LLM agents begin to offer epistemic value beyond automation. Fully agentic systems (Level 3) are increasingly used to model specific cognitive profiles, simulate participant roles, or replicate known psychological mechanisms \citep{liElicitingLanguageModel2025, staracePaperBenchEvaluatingAIs2025}. These agents support targeted experimentation and allow for controlled replications of empirical effects, strengthening trust in their validity \citep{karjusMachineassistedQuantitizingDesigns2025}. Multi-agent systems (Level 4) model social interactions by distributing capabilities across multiple autonomous entities. These configurations enable the study of inter-agent negotiation, group decision-making, norm formation, and institutional dynamics \citep{borghoffHumanArtificialInteractionAge2025,yangLLMMeasureGeneratingValid2024}. Particularly in political, organizational, or economic contexts, such systems facilitate experiments that would be logistically or ethically infeasible in real-world settings.

At Level 5, complex adaptive systems simulate population-scale phenomena. These models integrate thousands of interacting agents and allow for the exploration of emergent macro-social patterns, such as the spread of misinformation, social tipping points, or cultural diffusion \citep{hammondMultiAgentRisksAdvanced2025, parkGenerativeAgentsInteractive2023}. The ability to generate plausible yet novel outcomes---rather than merely reproduce existing ones---underscores the knowledge-generating potential of agentic simulations \citep{bailCanGenerativeAI2024, grossmannAITransformationSocial2023}. Anecdotally, the sophistication of current agentic systems is perhaps best illustrated by their deployment in complex gaming environments that require long-term strategic planning, resource management, and adaptive learning. Major technology companies have increasingly turned to Pokémon---a game requiring the training and strategic deployment of creatures in competitive battles---as a benchmark for testing advanced agentic capabilities. Anthropic's Claude 3.7 Sonnet achieved full game completion in February 2025 \citep{peterAnthropicsClaudeAI2025}, followed by Google AI's Gemini 2.5 in May of the same year \citep{schwartzGooglesGeminiAI2025}. These achievements demonstrate the capacity of contemporary LLM-based agents to engage in complex, multi-step decision-making processes that mirror the kind of strategic reasoning and adaptive behavior central to social coordination and collective problem-solving.

Across these tiers, LLM-based systems offer significant methodological gains. They enable scalable A/B testing, real-time iteration, longitudinal tracking without attrition, and high-throughput hypothesis evaluation. When validated against empirical benchmarks, these systems do not just replicate but extend the reach of social science methods, allowing researchers to model, test, and refine complex theoretical frameworks at a resolution previously inaccessible.

\subsection{Critical Reflections}

While the potential of LLM-based agents is considerable, their integration into social science raises serious methodological and ethical challenges. Chief among these is reproducibility. LLM outputs can vary across identical prompts due to stochastic generation and prompt sensitivity, especially in emergent, multi-agent settings where small changes can cascade into radically different outcomes \citep{atilLLMStabilityDetailed2024, luLLMsGenerativeAgentbased2024}. This complicates both experimental replication and the trustworthiness of emergent findings. The complexity of multi-agent systems also introduces new failure modes. As \citet{chanHarmsIncreasinglyAgentic2023} and \citet{hammondMultiAgentRisksAdvanced2025} show, agents may exhibit unintended behaviors such as conflict, miscoordination, or collusion, particularly when incentives or representations diverge. These behaviors are not merely technical bugs but epistemic distortions, especially if they arise from architectural choices rather than domain-relevant social processes.

Representation bias remains a critical issue. LLMs trained on predominantly Western, Anglophone corpora tend to underrepresent minority perspectives and may reinforce dominant cultural norms \citep{houNaturalLanguageProcessing2025, santurkarWhoseOpinionsLanguage2023}. This becomes especially problematic when agents are used as proxies for diverse human populations in public opinion research, policy modeling, or global simulations \citep{wangLargeLanguageModels2025, quPerformanceBiasesLarge2024}. Synthetic participants risk mischaracterizing the very populations they intend to model, introducing systemic distortions into research conclusions. In addition, LLMs lack many core elements of human cognition: genuine understanding, emotional depth, and meta-cognitive awareness \citep{rossiProblemsLLMgeneratedData2024,robertsArtificialIntelligenceQualitative2024}. Even when simulations are plausible at a surface level, they may fail to capture the mechanisms that drive real-world behavior. The assumption that linguistic mimicry equates to psychological realism must therefore be treated with caution.

Finally, the field suffers from a lack of methodological standardization. There is little consensus on how to evaluate agentic behavior, validate emergent phenomena, or benchmark performance across tasks \citep{keExploringFrontiersLLMs2024, luLLMsGenerativeAgentbased2024}. Further, especially for complex adaptive systems, there is no baseline to what to compare emergent phenomena to. While this lack of standardization may be understandable given that the field is relatively new and emerging, methodological standards need not await decades of development---they can be established by building upon the extensive methodological learning curves and validation frameworks that social science has already developed over its long history. This hinders scientific progress, limits comparability across studies, and raises barriers to reproducibility. Without shared standards, the promise of LLM agents risks being diluted by inconsistent or poorly justified findings.

Taken together, these challenges should not be mistaken as disqualifying the use of agentic systems in social science. Rather, they highlight the necessity of cautious, rigorous, and interdisciplinary development. Like traditional social science methods, which have long grappled with the tension between representing complex social realities and simplifying them to identify causal mechanisms \citep{mwitaStrengthsWeaknessesQualitative2022, hofmanIntegratingExplanationPrediction2021}, the use of LLM-based agents requires a similar balance. The limitations of agentic systems---be they related to reproducibility, bias, or representational fidelity---mirror long-standing critiques of social science itself, which is often accused of being too reductive or too speculative. Yet, despite these critiques, decades of careful, cumulative, and reflective research have demonstrated the field's ability to produce meaningful insights \citep{milleEffectsContinuousDiscontinuous2022}. Similarly, when embedded within robust research designs, subjected to critical validation, and aligned with a culture of transparency and replication, LLM-based agents can offer significant epistemic value \citep{freeseReplicationSocialScience2017}. With appropriate safeguards and theoretically informed constraints, they are not just technical tools, but potential contributors to a cumulative science of social behavior.

\subsection{Future Research Directions}

The rapid development of LLM-based agents presents a wide range of opportunities for advancing social science research. However, realizing this potential will require both technical innovation and interdisciplinary collaboration. The following directions highlight key areas where future work can contribute to the maturation of this field:

\paragraph{Advancing Agent Capabilities.} Future research should focus on enhancing the cognitive and emotional sophistication of LLM-based agents. This includes integrating memory mechanisms, affective reasoning, and goal-oriented planning to create more human-like, context-sensitive agents capable of long-term stable interaction. Advanced architectures, such as Centaurian systems, which combine the strengths of human cognition and machine processing, offer promising paths forward \citep{borghoffHumanArtificialInteractionAge2025}. Additionally, further exploration is needed to understand how agentic systems can simulate complex social processes, including identity formation, social learning, and collective problem-solving, without falling into the traps of oversimplification common for traditional agentic systems \citep{loweMultiAgentActorCriticMixed2017}.

\paragraph{Methodological Innovations.} Given the inherent complexity of multi-agent systems, robust validation techniques are essential. This includes the development of standardized evaluation metrics, reproducibility protocols, and benchmarking strategies that capture both the intended behaviors and the unintended emergent properties of these systems. Recent work has highlighted the need for more consistent and interpretable outputs in LLM-based simulations, particularly in light of known stability issues \citep{atilLLMStabilityDetailed2024}. As these systems move from experimental to applied settings, the ability to ensure reliability, transparency, and ethical consistency will become increasingly critical. This also includes addressing the inherent biases in language models, which can propagate into social simulations if left unchecked \citep{keExploringFrontiersLLMs2024}.

\paragraph{Expanding Application Domains.} The flexibility of LLM-based agents offers unique opportunities for extending their use beyond traditional computational settings. Future research should explore cross-cultural, multilingual, and real-world decision-making contexts to capture the full diversity of human social behavior. This could involve integrating LLMs into real-time, high-stakes decision-making processes, such as crisis response, policy simulation, or economic forecasting, where their ability to rapidly process and synthesize large volumes of data can provide significant advantages \citep{murakamiMultiagentSimulationCrisis2002}. Additionally, multi-agent systems present a valuable opportunity for studying the dynamics of global phenomena, such as misinformation spread, political polarization, and international collaboration, under controlled but realistic conditions.

\paragraph{Human-AI Collaboration and Hybrid Systems.} As agentic systems become more capable, their integration into human workflows across all stages of the research process will require careful consideration. This includes not only technical design but also the creation of ethical guidelines, transparency standards, and training protocols that ensure humans remain in control of critical decision-making processes (e.g., for ChatGPT and scientific authorship see \citet{editorialAIWritingWall2023}). Recent research has highlighted the need for hybrid systems that effectively integrate human intuition with machine-scale processing, creating a balanced division of cognitive labor that leverages the strengths of both \citep{borghoffHumanArtificialInteractionAge2025,haaseHumanAICoCreativityExploring2024}. This approach could significantly enhance the practical applicability of agentic systems, transforming them from experimental tools into robust components of real-world decision support systems.

\paragraph{Fundamental Theoretical Questions.} Finally, researchers should address the deeper theoretical implications of using LLMs as proxies for human behavior. This includes examining the philosophical and ethical dimensions of substituting human participants with synthetic agents, as well as the epistemological risks associated with treating LLM outputs as stand-ins for human cognition \citep{zhouThisRealLife2024, larooijLargeLanguageModels2025}. Critical reflections are needed to ensure that these systems genuinely advance social science rather than merely replicate its methods and challenges at scale. This requires ongoing dialogue between computer scientists, social scientists, ethicists, and policymakers to establish clear guidelines for the responsible use of agentic systems in research and practice.

\section{Conclusion}

This paper proposes a structured framework for integrating LLM-based agents into social science research, outlining a six-tier model that spans from simple stateless tools to fully adaptive multi-agent systems. This continuum offers a conceptual and practical foundation for both researchers and system designers to assess the capabilities, requirements, and epistemic roles of LLM agents in social inquiry. At the lower tiers, agentic systems offer clear methodological benefits: streamlining repetitive tasks, improving reproducibility, and enabling scalable text analysis. At the higher levels, however, the value of LLM agents shifts from efficiency toward epistemic innovation. Here, multi-agent architectures simulate emergent social behavior, test theoretical propositions at scale, and model phenomena, such as norm formation, conflict dynamics, or policy effects, that are difficult, if not impossible, to study through traditional empirical means. These systems, when properly constrained and evaluated, offer genuinely novel ways of generating social scientific insight.

Still, this promise is not without its caveats. The challenges of reproducibility, representational bias, and epistemological overreach demand careful, interdisciplinary scrutiny. Yet, these issues are not unique to agentic systems---they echo long-standing tensions in the social sciences themselves. As with any simplification of complex human phenomena, the power of LLM-based agents lies not in perfect replication but in their capacity to isolate, model, and explore meaningful mechanisms under controlled conditions. When embedded within transparent research designs, aligned with cumulative scientific standards, and subjected to replication and critique, agentic systems can meaningfully contribute to social scientific knowledge.

Looking forward, key priorities for future work include expanding agent capabilities (e.g., memory, affective reasoning, long-term goal adaptation), refining standards for robustness and validation, and applying these systems across a broader range of cultural and institutional contexts. In parallel, their integration into human workflows must be guided by ethical design principles and supported by interdisciplinary collaboration. Together, these efforts chart a path toward a more integrated, responsible, and impactful role for LLM-based agents in the social sciences. Rather than replacing traditional methods, these systems augment and extend them---opening up new ways of seeing, modeling, and understanding the complex dynamics that shape human societies.

\section{Acknowledgments}

Research reported in this paper was partially supported by the Deutsche Forschungsgemeinschaft (DFG) through the DFG Cluster of Excellence MATH+ (grant number EXC-2046/1, project ID 390685689), and by the German Federal Ministry of Education and Research (BMBF), grant number 16DII133 (Weizenbaum-Institute).

\bibliographystyle{plainnat}
\bibliography{references,references_shared_BIB,extra}

\clearpage

\appendix

\end{document}